\documentclass[twocolumn,showpacs]{revtex4}
\usepackage[xdvi]{graphicx}%
\usepackage{dcolumn}
\usepackage{amsmath}
\makeatletter

\def\btt#1{\texttt{\@backslashchar#1}}%
\DeclareRobustCommand\bblash{\btt{\@backslashchar}}%
\makeatother

\newcommand{\bra}{\left\langle}
\newcommand{\ket}{\right\rangle}
\newcommand{\der}[2]{\frac{d #1}{d  #2}}
\newcommand{\pder}[2]{\frac{\partial #1}{\partial  #2}}
\newcommand{\pdert}[2]{\frac{\partial^2 #1}{\partial  #2^2}}

\newcommand{\ps}{p_{\rm s}}

\newcommand{\vs}{v_{\rm s}}

\newcommand{\fm}{f_{\rm m}}
\newcommand{\ca}{{\cal A}}
\newcommand{\cb}{{\cal B}}
\newcommand{\cc}{{\cal C}}
\newcommand{\ve}{\varepsilon}
\newcommand{\e}{{\rm e}}
\newcommand{\mud}{\mu_{\rm d}}

\begin{document}

\title{Effective temperature in nonequilibrium steady states of Langevin 
systems with a tilted periodic potential}
\author{Kumiko Hayashi and Shin-ichi Sasa} 
\affiliation
{Department of Pure and Applied Sciences,
University of Tokyo, Komaba, Tokyo 153-8902, Japan}

\date{\today}

\begin{abstract} 
We  theoretically study  Langevin systems with a tilted periodic potential. 
It has been known that the ratio $\Theta$ of the diffusion constant $D$ to the 
differential mobility $\mud$ is not equal to the temperature of the 
environment (multiplied by the Boltzmann constant), except in the linear 
response regime, where the fluctuation dissipation theorem holds.
In order to elucidate the physical meaning of $\Theta$ far from equilibrium, 
we analyze a modulated system with a slowly varying potential. 
We derive a large scale description of the probability density
for the modulated system by use of  a perturbation method. 
The expressions we obtain show that $\Theta$ plays the role of the temperature 
in the large scale description of the system and that $\Theta$ can 
be determined directly in experiments, without measurements of the diffusion 
constant and the differential mobility. 
Hence the relation $D=\mud\Theta$  among independent measurable 
quantities $D$, $\mud$ and $\Theta$ can be interpreted as 
an extension of the Einstein relation.
\end{abstract}

\pacs{05.40.-a, 02.50.Ey, 05.70.Ln}
\maketitle

\section{Introduction}

Technological development in both the manipulation  and 
observation of objects on small scales has led to a new understanding
of the behavior exhibited by mechanical systems of length scales 
$\sim 10^{-6}$ m, force scales $\sim  10^{-12}$ N, and the time scales
$\sim 10^{-3}$ sec. One such system that has been studied consists of
a small bead suspended in a fluid.  Because of their simplicity, such 
systems are particularly useful for studying topics relevant to fundamental
physics. Indeed, 
recently, the fluctuation theorem \cite{FT} and the Jarzynski equality 
\cite{jar},
which were   derived theoretically as  relations universally valid 
for nonequilibrium processes, have been verified experimentally 
through experiments on systems consisting of small beads \cite{ftex} and 
RNA molecules \cite{jarex} employing optical tweezers. 

  
In studying nonequilibrium systems,  we want to discover the uniquely 
nonequilibrium behavior as well as to determine what
properties of equilibrium systems remain even far from equilibrium.
We believe that small systems are suited for such studies because 
nonequilibrium effects become more  significant  as the system
size decreases. In particular, with regard to nonequilibrium systems, 
we are interested in finding  new  relations between measurable 
quantities that may be useful in the construction of a systematic 
theory of  nonequilibrium statistical mechanics.


In the present paper, we study the motion of a small bead  suspended in a 
fluid of temperature $T$. The bead  is confined to move in a single direction, 
say the $x$-direction, and is subjected to a periodic potential  $U(x)$ 
of period $\ell$. Such a system can be realized experimentally as 
a scanning optical trap system \cite{hara}, for example.  Further, 
a flow with constant velocity can be used to apply a constant driving force 
$f$ to the bead.  In this way, it is possible to experimentally realize
nonequilibrium steady states (NESS) for such a bead system.


The quantity we investigate  is the ratio $\Theta$ 
of the diffusion constant $D$ to the differential mobility $\mud$ 
for the bead in NESS.  In the linear response regime, the ratio 
$\Theta$ is identical to the temperature of the environment (multiplied by 
the Boltzmann constant). This relation is equivalent to one form of
the fluctuation dissipation theorem (FDT). However, because the FDT
does not hold for NESS far from equilibrium, $\Theta$ is not identical 
to the temperature of the environment.  Nevertheless, in the system 
considered there, we find that $\Theta$ plays the role of the 
temperature in the description of the large scale behavior of 
the system and that $\Theta$ can be determined experimentally
in a direct manner, without the need to measure $D$ and $\mud$. 
We  obtain this result by employing  a perturbation method 
to derive the large scale description of the probability density.

\section{Model}

We assume that motion of the bead is described by the one-dimensional 
Langevin equation  
\begin{equation}
\gamma\dot{x}=-\frac{\partial U(x)}{\partial x}+f
+\sqrt{2\gamma T}\xi(t),
\label{lan} 
\end{equation}
where $\xi(t)$ is a Gaussian white noise with zero mean and unit 
dispersion.  The Boltzmann constant is set to unity.  We note
that there are many physical examples that are described by this
equation \cite{tpp}.  
Here, we make two remarks on the form of
(\ref{lan}).  First, inertial effects are considered to be negligible, 
because  a typical value of the relaxation time of the particle velocity,
which  is estimated to be $\sim 10^{-9}$ sec \cite{est}, is much shorter 
than the characteristic time scale of the phenomena observed in the type 
of experiments we consider. Second, $T$ in (\ref{lan}) is assumed to be 
the temperature of the environment. Although the validity of this assumption 
is not known for NESS in general, our result does not depend on the 
physical interpretation of $T$, because we do not need to use the 
value of $T$ in our analysis.


The probability density for the position of the particle $p(x,t)$ 
in this system obeys the Fokker-Planck equation 
\begin{equation}
\pder{p}{t}=\frac{1}{\gamma}\pder{}{x}
\left[ \left( \pder{U}{x}-f \right) p+T\pder{p}{x} \right].
\label{fp}
\end{equation}
The steady state density $\ps(x;f)$ is obtained  as \cite{Risken}
\begin{equation}
\ps(x;f)=\frac{1}{Z} I_-(x)
\label{ps}
\end{equation}
with
\begin{equation}
I_-(x)=\int_0^\ell dy \e^{-\beta U(x)+\beta U(y+x)-\beta fy},
\end{equation}
where $\beta=1/T$ and  $Z$ is a normalization factor by which   
\begin{equation}
\int_0^{\ell}dx\ps(x;f)=\ell.
\end{equation}
The steady state current $\vs(f)$ is derived as
\begin{equation}
\vs(f)=\frac{T}{\gamma}\frac{1-\e^{-\beta f \ell}}
{\frac{1}{\ell}\int_0^\ell dx I_-(x)}.
\label{cur}
\end{equation}
The exact expression of the diffusion constant $D(f)$ has been 
derived recently for the system under consideration \cite{RH}. 
By using a different method (See sections \ref{s:an} and \ref{s:te}), 
we derive the following form for $D(f)$:
\begin{equation}
D(f)=\frac{T}{\gamma} 
\frac{\frac{1}{\ell}\int_0^\ell dx (I_-(x))^2 I_+(x)}
{\left(\frac{1}{\ell}\int_0^\ell dx I_-(x) \right)^3},
\label{dif}
\end{equation}
where 
\begin{equation}
I_+(x)=\int_0^\ell dy e^{\beta U(x)-\beta U(x-y)-\beta fy}.
\end{equation}
Note that (\ref{dif}) is equivalent to the expression derived in Ref. 
\cite{RH}, which is obtained by simply exchanging $I_+$  and $I_-$ 
in (\ref{dif}).

\begin{figure}
\begin{center}
\includegraphics[width=8cm]{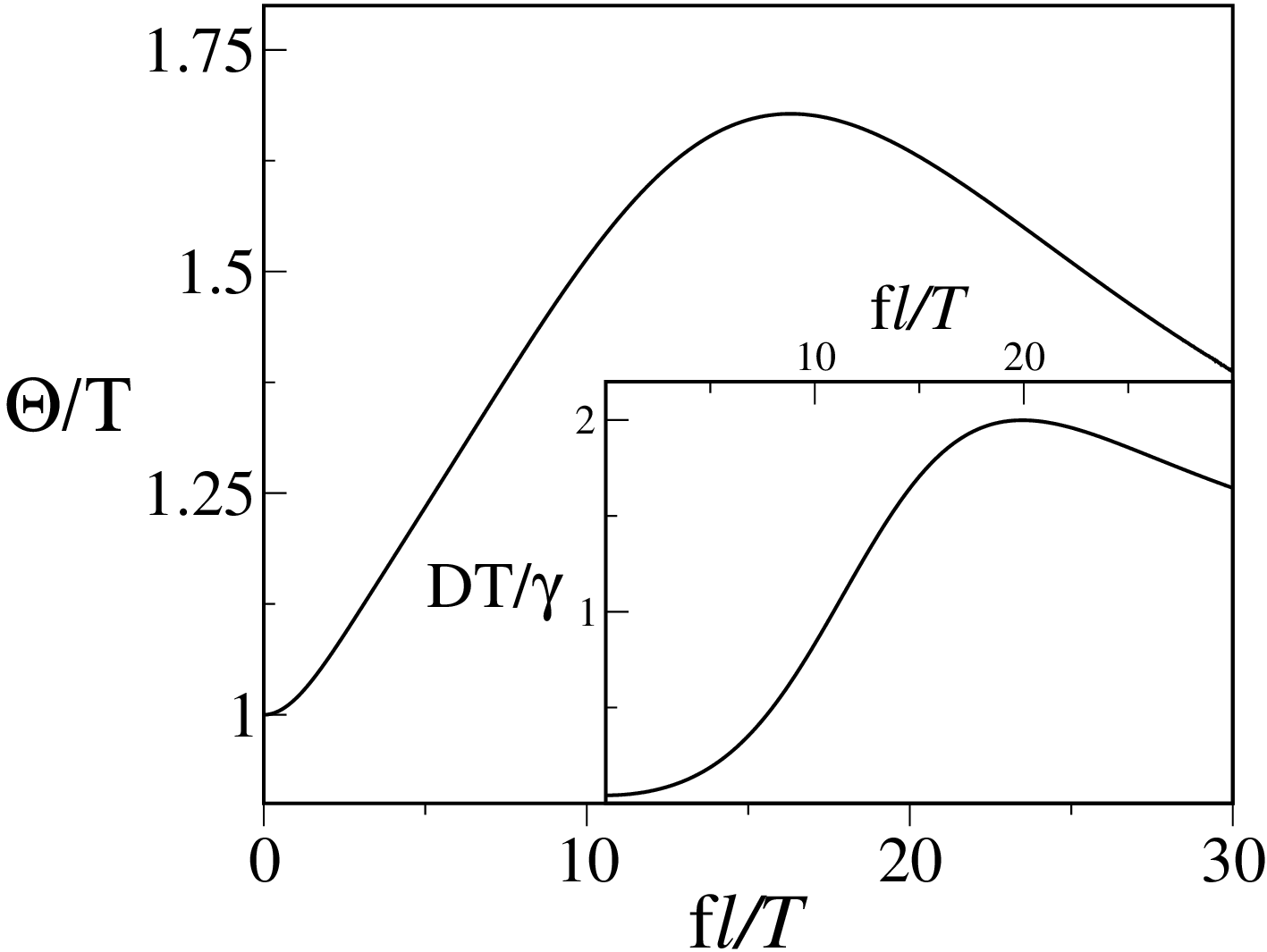}
\end{center}
\caption{$\Theta/T$ as a function of $f\ell/T$ in the case that
$U(x)=U_0\sin2\pi x/\ell$ with $U_0/T=3.0$. $\Theta/T=1$ at $f=0$, 
and $\Theta/T > 1$ for $f >0$. 
The inset displays $D\gamma/T$ as a function of $f\ell/T$ calculated 
using (\ref{dif}).}
\label{temp}
\end{figure}

In the inset of Fig. \ref{temp}, we display an example of $D(f)$ for 
the case  $U(x)=U_0 \sin 2 \pi x/\ell$. It can be seen that the 
diffusion is enhanced 
around $f\ell/T= 2 \pi U_0/T$. This effect was first reported 
in Ref. \cite{CM} and was subsequently analyzed more 
quantitatively \cite{RH}.

\section{Question}

In the linear response regime, the mobility $\mu$, defined as
$\mu= \lim_{f \to 0}\vs(f)/f$, is equal to  $D(f=0) T$. This 
is one form of the FDT. However, for NESS far from equilibrium, 
the FDT does not hold. In fact, far from equilibrium, there does
not even exist an FDT involving the differential mobility 
$\mud \equiv d\vs/df$, drived from (\ref{cur}) as follows: 
\begin{equation}
\mud(f)=\frac{1}{\gamma} 
\frac{\frac{1}{\ell}\int_0^\ell dx I_-(x)I_+(x)}
{\left(\frac{1}{\ell}\int_0^\ell dx I_-(x) \right)^2}.
\label{dmob}
\end{equation}

In order to determine quantitatively the extent to which the 
FDT is violated, we define $\Theta$ as the ratio of $D$ to $\mud$:
\begin{equation}
\Theta(f) \equiv \frac{D(f)}{\mud(f)}.
\label{fdt2}
\end{equation}
As displayed in Fig. \ref{temp}, although  the dimensionless quantity  
$\Theta/T$ is unity in equilibrium, it depends on $f \ell/T$ 
and on the form of the periodic potential $U(x)$ when the system is
far from equilibrium. 
In addition to the fact that the FDT does not hold far
from equilibrium, no means of measuring $\Theta$ experimentally without
measuring $D$ and $\mud$ has  been proposed. Hence,  
there is no known method of deducing $\mud$ from experimentally
measured values of $D$. 
While the  fluctuation theorem  
reported in Ref. \cite{FT}, which is valid for NESS far from equilibrium,
can be regarded as an extension of FDT \cite{Gall}, it seems difficult to 
connect it to a physical interpretation of (\ref{fdt2}).

Despite the apparent difficulties described above, we attempt to
propose a method of measuring $\Theta$ experimentally without 
measuring $D$ and $\mud$. 
We first note that $\Theta/T$ ($\ne 1$ in NESS) is interpreted as the 
FDT violation factor.
Recently, stimulated by a  proposal of the thermodynamic measurement of  
the FDT violation factor in spin glass systems  as  the ''effective 
temperature'' \cite{CKP}, the feasibility of physical measurements 
of the effective temperature has been investigated in numerical experiments 
modeling  a sheared glassy material \cite{BB}, a driving system near 
jamming \cite{jamm}, and driven vortex lattices \cite{DVL}.  Among them, 
Berthier and Barrat have proposed a novel measurement method of the 
FDT violation factor as the effective temperature in shared glassy
systems \cite{BB}.  They have used a tracer particle of large mass
as an effective thermometer and demonstrated that the kinetic 
energy of the tracer is related to the FDT violation factor. 
This result exhibits a clear 
relation between the FDT violation factor and the measurable effective 
temperature. 

Here too we seek to determine whether there is a realm in which 
$\Theta$ can be interpreted as the effective temperature 
in NESS and, hence, whether it can be measured independently
of $D$ and $\mud$.

\section{Result}\label{s:an}

We propose a method of measuring $\Theta$  by adding a slowly varying 
potential which plays the role of an effective thermometer to the system 
under consideration. 
In order to demonstrate this method, we  
study (\ref{lan}) with $U(x)$ replaced by a potential 
$U(x)+V(x)$, where $V(x)$ is  a slowly varying periodic 
potential $V(x)$ of period  $L \gg \ell$. We investigate the
behavior of the probability density on length scales larger than $L$. 
In order to make the separation of scales explicit, we define 
$\ve \equiv \ell/L$ and the large scaled coordinate $X \equiv\ve x$. 
The probability density in this system obeys the  Fokker-Planck equation
\begin{equation}
\pder{p}{t}=\frac{1}{\gamma}\pder{}{x}
\left[ \left(\pder{U}{x}+\ve\pder{\tilde V}{X}-f \right) p
+T\pder{p}{x} \right],
\label{fpm}
\end{equation}
where we have defined $\tilde V(X) \equiv V(x)$. 

We extract the large scale behavior of $p(x,t)$ by introducing
a slowly varying field $Q(X,t)$ as 
\begin{eqnarray}
p(x,t)&=& \ps(x;\fm)\left(
Q+ \ve a(x;\fm) \pder{Q}{X}
+ \ve^2 b(x;\fm) \pdert{Q}{X} \right. \nonumber \\
&+& \left.
\ve c(x;\fm) \pder{\fm}{X} Q+O(\ve^3) \right), 
\label{defQ2}
\end{eqnarray}
where  $\fm=f-\ve \partial \tilde V/\partial X$. Because $\fm$ is 
a function of 
$X=\ve x$, the $x$ dependence of $\ps(x;\fm)$ appears in the two ways,
as an explicit dependence  and as a dependence through $\fm$.
Note that $\ps(x;\fm)$ is a periodic function in the sense that 
$\ps(x+\ell;\fm)=\ps(x;\fm)$. The functions $a(x;\fm)$, $b(x;\fm)$ and 
$c(x;\fm)$ are similar; that is, they are periodic functions  of $x$ 
in the same sense and depend on $\fm$. 
Their functional forms are determined below.  

\begin{figure}
\begin{center}
\includegraphics[width=8cm]{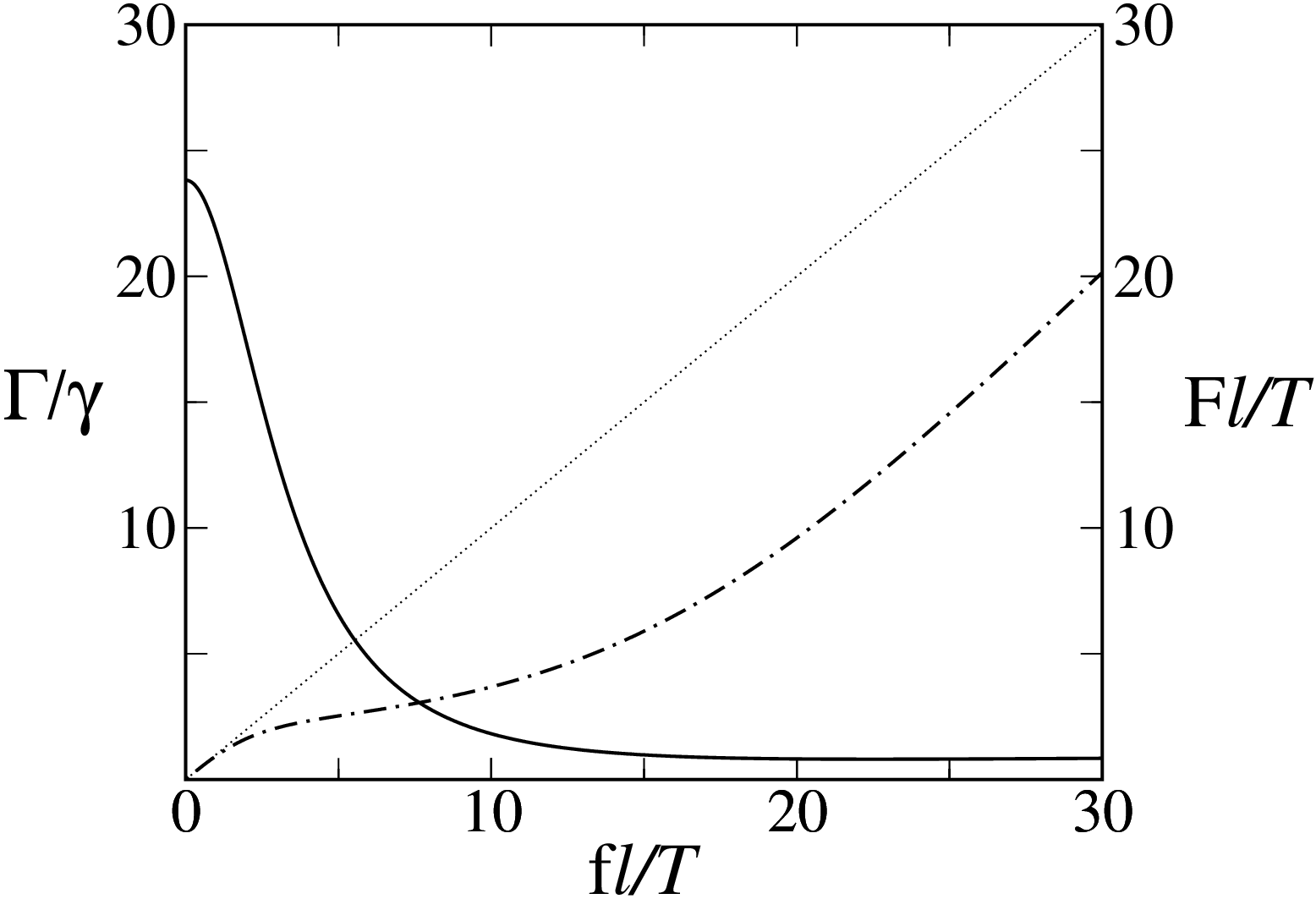}
\end{center}
\caption{The left axis represents $\Gamma/\gamma$, which is displayed as 
a function of $f\ell/T$ (solid curve), and the right axis represents 
$F\ell/T$, also displayed as a function of $f\ell/T$ (dash-dotted curve), 
in the case that $U(x)=U_0\sin2\pi x/\ell$, with $U_0/T=3.0$.  The 
dotted line corresponds to $F=f$. It is seen that in the linear response 
regime, the effective force $F$ is equal to $f$.  
}
\label{force}
\end{figure}

Substituting (\ref{defQ2}) into (\ref{fpm}), we obtain
\begin{equation}
\pder{Q}{t}= \ve \ca \pder{Q}{X}+\ve^2 \cb \pdert{Q}{X}
+\ve \cc \pder{\fm}{X}Q +O(\ve^3),
\label{nf22}
\end{equation}
where $\ca$, $\cb$ and $\cc$ can be expressed by
\begin{eqnarray}
\ca &=& -\left(\vs -\frac{T}{\gamma} {\ps}^\prime \right)\ps^{-1}
(1+a^\prime)+\frac{T}{\gamma}a^{\prime\prime}, \label{ca} \\
\cb &=& -\left(\vs -\frac{T}{\gamma} {\ps}^\prime \right)\ps^{-1}
(a+b^\prime) \nonumber \\
&+ &\frac{T}{\gamma}(1+2a^\prime+b^{\prime\prime})
-a\ca, \label{cb}\\
\cc &=&  -\left(\vs -\frac{T}{\gamma} {\ps}^\prime \right)\ps^{-1}
c^\prime+\frac{T}{\gamma}c^{\prime\prime}  \nonumber \\
&-& \der{\vs}{\fm}\ps^{-1}+\frac{T}{\gamma}\pder{\ps^\prime}{\fm}\ps^{-1},
\label{cc}
\end{eqnarray}
where ${}^\prime$ represents the partial derivative with respect to $x$,
that is, $a'(x;\fm)=\partial a(x;\fm)/\partial x$ and so on.
Then we can choose $a$, $b$ and $c$ so that  $\ca$, 
$\cb$ and $\cc$ do not depend on $x$ explicitly but depend on $x$ 
through the $X$ dependence of $\fm$. After a straightforward calculation, 
we find that, subject to this condition, $\ca$, $\cb$ and $\cc$ are uniquely 
determined as 
\begin{eqnarray}
\ca &=& -\vs(\fm),
\label{c2a} \\
\cb &=& D(\fm),
\label{c2b} \\
\cc &=&  -\der{\vs(\fm)}{\fm}.
\label{c2c}
\end{eqnarray}
(The derivation  will be presented in section \ref{s:te}.) 
Using this result,  we rewrite (\ref{nf22}) as
\begin{equation}
\pder{Q}{t}= \ve\pder{}{X}\left[
 -\vs(\fm) Q + D(\fm) \ve\pder{Q}{X}
+O(\ve^2) \right].
\label{nf23}
\end{equation}
Recalling that $\fm=f-\ve \partial \tilde V/\partial X$, we see that 
this equation is equivalent to the Fokker-Planck  equation
\begin{equation}
\pder{Q}{t}= \frac{1}{\Gamma}\ve\pder{}{X}\left[
\left(\ve\pder{\tilde V}{X}-F\right) Q+ \Theta \ve \pder{Q}{X}
+O(\ve^2) \right],
\label{FP}
\end{equation}
where we have defined 
\begin{eqnarray}
\Gamma(f) &\equiv&  \mud(f)^{-1}, \label{gdef} \\
F(f) &\equiv& \vs(f) \mud(f)^{-1}. 
\end{eqnarray}
For reference, in Fig. \ref{force}, we present
graphs of $\Gamma$ and $F$ as a function of $f$ for the model 
considered in Fig. \ref{temp}.

We note that (\ref{FP}) has the same form as (\ref{fp}), with
the parameters $\Gamma$, $F$ and $\Theta$ corresponding to  $\gamma$, $f$ 
and $T$, respectively. Thus, replacing ($\gamma$,$f$,$T$,$U(x)$) 
in  (\ref{cur}) by ($\Gamma$,$F$,$\Theta$,$\tilde V(X)$), 
we obtain  the current for NESS of this modulated 
system. {}From this result, by measuring the steady state 
current for several forms of $ V(x)$, we can determine the values
of $\Gamma$, $F$ and $\Theta$ experimentally, 
where we note that these values  do not 
depend on the choice of $V(x)$.  
This fact implies the existence of the new relation  $D=\mud\Theta$  
among the independent measurable quantities $D$, 
$\mu_d$ and $\Theta$ in NESS far from equilibrium.    
 This  new relation $D=\mud\Theta$  can be thought as the extended 
Einstein relation. 
Furthermore, from the correspondence between  $\Theta$ in (\ref{FP}) and 
$T$ in (\ref{fp}), it is evident  that $\Theta$ plays the role of the 
temperature for the large scale behavior of the system.
In this way, we have arrived at the main claim of this paper, the physical 
interpretation of $\Theta$.  

Here we address two remarks on the main claim. 
First, one may naively expect that  on the large scale (\ref{lan}) can be 
effectively described by 
\begin{equation}
\dot{x}=\vs(f)+\sqrt{2D(f)}\xi(t).
\label{lan2} 
\end{equation}
Although such an effective description is valid, we emphasize that
the effective temperature is not determined from this description.
In order to have the correspondence with (\ref{FP}), we need one 
more quantity in addition to $D$ and $\vs$.  In our analysis, by
adding the slowly varying potential to the system, $\Gamma$, $F$ 
and $\Theta$ are determined. We thus interpret the slowly varying 
potential as an effective thermometer for the original system 
described by  (\ref{lan}). 

Second, we  have another method of measuring $\Theta$ by  considering 
the case in which  a slowly varying potential 
$\tilde V(\ve(x-\vs t))$ that moves with  constant velocity $\vs$ 
replaces $\tilde V(\ve x)$ in (\ref{fpm}). Although the resulting system 
is more complicated than that considered above, the large scale 
behavior in this case is actually simpler. Indeed, using the same 
method as above, we obtain the following equation for $Q(Y,t)$ 
describing the large scale  behavior of the system:
\begin{equation}
\pder{Q(Y,t)}{t}= \frac{1}{\Gamma}\ve \pder{}{Y}
\left (\ve\pder{\tilde V(Y)}{Y} Q +  \Theta \ve \pder{Q}{Y} +O(\ve^2) \right).
\label{FP2}
\end{equation}
Here, we have introduced the large scaled moving coordinate 
$Y\equiv \ve(x-\vs t)$. This equation is identical to the 
Fokker-Planck equation describing the time evolution of 
the probability density in an equilibrium state with 
temperature $\Theta$.  Therefore, for example, $\Theta$ 
is obtained  by measuring the statistical average of $\tilde V(Y)$.

\section{numerical experiment}

\begin{figure}
\begin{center}
\includegraphics[width=8cm]{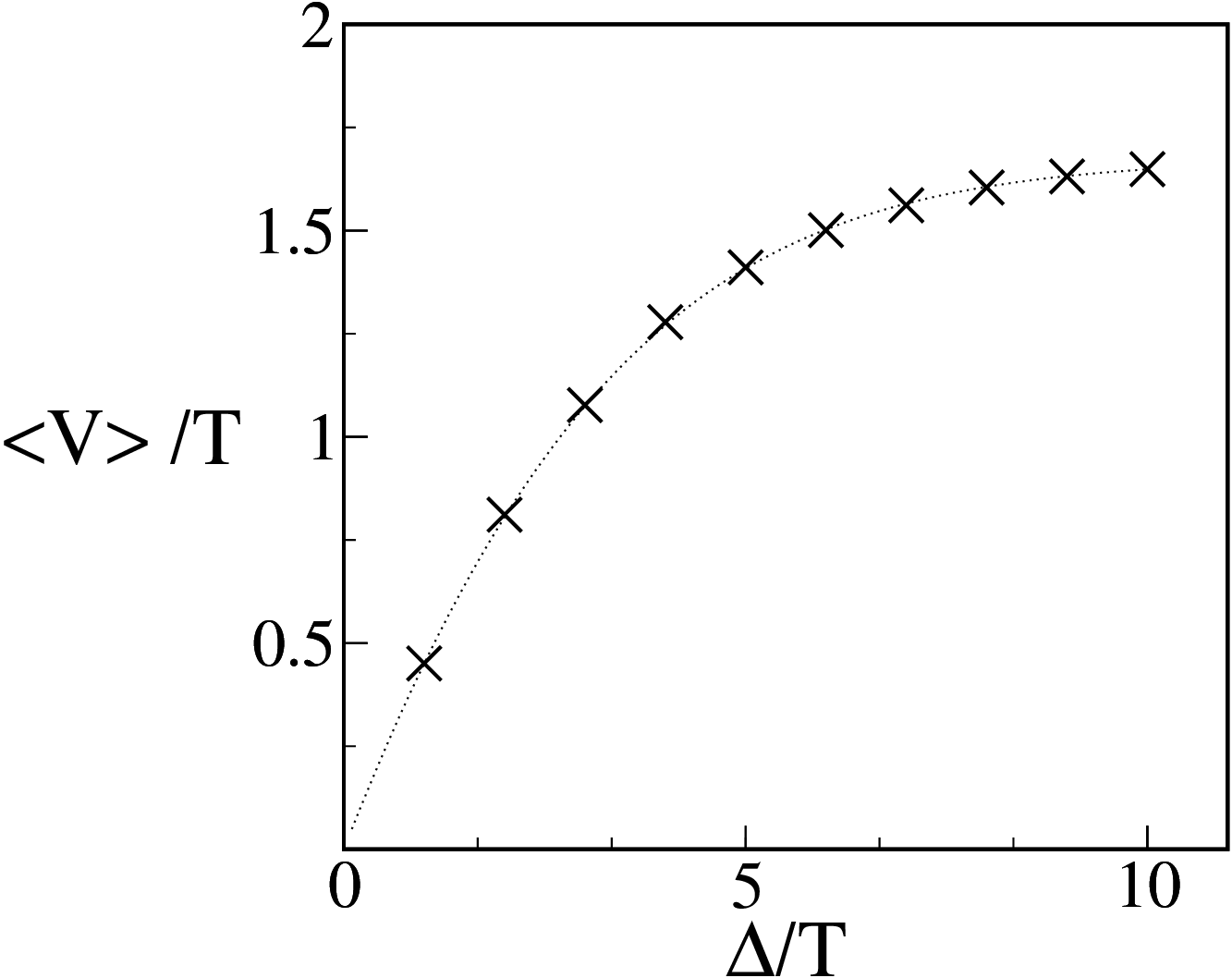}
\end{center}
\caption{
$\bra V \ket/T$ versus  $\Delta/T$ obtained in the numerical experiment. 
The dotted line represents the fitting curve of the form (\ref{form}) 
in which $\Theta/T$=1.67(4).  
}
\label{vdelta}
\end{figure}

Let us demonstrate that the measurement method of $\Theta$ 
by use of a slowly varying potential works well in numerical 
experiments. We study numerically (\ref{lan}) with periodic 
boundary conditions for the case that $U(x)=U_0 \sin 2 \pi x/\ell$ 
and that the system size is $L$. The parameter values are chosen 
as follows: $L/\ell=50$, $U_0/T=3$ and $f\ell/T= 16$. 

We measure the effective temperature for this system by using
a moving potential $V(y)$ in the form
\begin{eqnarray}
V(y) &=& 2\Delta y /L  \quad {\rm for} \quad 0 \le y \le L/2 \nonumber \\
        &=& 2 \Delta-2\Delta y /L  \quad {\rm for} \quad L/2  \le y \le L,
\end{eqnarray}
with $y={\rm mod}(x-\vs t, L) $. The statistical averages of $V$ in NESS,
$\bra V \ket$, are measured for several values of $\Delta$. As displayed
in  Fig. \ref{vdelta}, the graph $(\Delta, \bra V \ket)$ is fitted well 
to the form 
\begin{equation}
\bra V \ket=\Theta - \frac{\Delta}{\exp(\Delta/\Theta)-1},
\label{form}
\end{equation}
which can be expected from (\ref{FP2}). Using this fitting for 
the experimental result, we evaluate the value of $\Theta/T$  
as $\Theta/T=1.67(4)$.   
This value should be compared with the theoretical one, 1.677, 
which is calculated from  (\ref{fdt2}). Thus we conclude that  
our measurement method of $\Theta$ works well in experiments.

\section{technical details}\label{s:te}

We regard (\ref{ca}) as an ordinary differential equation (ODE) for
$a(x;\fm)$ under the condition that $\ca$ does not depend on $x$. 
This ODE has periodic solutions of $x$ only when $\ca$ satisfies 
a certain condition,  which  provides uniquely $\ca$. 

Let us solve the ODE (\ref{ca}). We first define  
\begin{equation}
d(x;\fm)=a'(x;\fm)\ps(x;\fm).
\end{equation}
Then (\ref{ca}) becomes
\begin{equation}
\frac{T}{\gamma}d^{\prime}-\vs a^\prime -\vs+\frac{T}{\gamma}\ps^{\prime}
-\ca \ps=0.
\label{ca2}
\end{equation}
Integrating this equation over the range $[0, \ell]$, we obtain
\begin{equation}
\ca=-\vs,
\label{caa}
\end{equation}
where we have used a requirement that $Td/\gamma-\vs a$ is
a periodic function.

Under the condition (\ref{caa}), we can derive periodic solutions $a(x;\fm)$. 
The integration of (\ref{ca2}) leads 
\begin{equation}
\frac{T}{\gamma}a'\ps -\vs a -\vs x+\frac{T}{\gamma}\ps+\vs H +\vs K_1=0,
\label{ca3}
\end{equation}
where $K_1$ is a constant and we have defined 
\begin{equation}
H(x;\fm)=\int_0^x dy \ps(y;\fm).
\end{equation}
Substituting  the expression
\begin{equation}
a(x;\fm)=H(x;\fm)-x+\bar a(x;\fm)
\label{abar}
\end{equation}
into (\ref{ca3}), we rewrite (\ref{ca3}) as 
\begin{equation}
\frac{T}{\gamma}\ps \bar a'-\vs \bar a +\frac{T}{\gamma}\ps^2+\vs K_1=0.
\label{ca4}
\end{equation}
Noting the relation
\begin{equation}
\vs \left(\ps \e^{\beta (U(x)-\fm x)}\right)
=-\frac{T}{\gamma}\ps  \left(\ps \e^{\beta (U(x)-\fm x)}\right)^\prime,
\end{equation}
we  solve (\ref{ca4}) as 
\begin{eqnarray}
\bar a(x;\fm) &=& -\ps(x;\fm)^{-1} \e^{-\beta (U(x)-\fm x)}\Phi(x;\fm) 
\nonumber \\
 &+& K_2 \ps(x;\fm)^{-1} \e^{-\beta (U(x)-\fm x)}+K_1,
\label{abar2}
\end{eqnarray}
where we have defined 
\begin{equation}
\Phi(x;\fm)=\int_0^x dy \ps(y;\fm)^2\e^{\beta (U(y;\fm)-\fm y)}.
\label{phi}
\end{equation}
{}From the condition $a(0;\fm)=a(\ell;\fm)$, the constant $K_2$ is
determined as
\begin{equation}
K_2=\frac{1}{1-\e^{-\beta \fm \ell}}\Phi(\ell;\fm).
\label{k2}
\end{equation}
(\ref{abar}), (\ref{abar2}), (\ref{phi}) and (\ref{k2}) provide
all periodic solutions of ODE (\ref{ca}).

Next we study (\ref{cb}) by repeating the similar analysis.
Defining 
\begin{equation}
h(x;\fm)=b'(x;\fm)\ps(x;\fm),
\end{equation}
we rewrite  (\ref{cb}) as
\begin{equation}
\frac{T}{\gamma}h^{\prime}-\vs b^\prime 
-\vs a
+\frac{T}{\gamma}(1+2a') \ps
+\frac{T}{\gamma}a \ps^{\prime}
-\cb \ps
+a \vs \ps=0.
\label{cb2}
\end{equation}
Integrating this equation over the range $[0, \ell]$, we obtain
\begin{equation}
\cb=\frac{T}{\gamma}+\frac{T}{\gamma \ell}\int_0^\ell dx a' \ps
-\frac{\vs}{\ell}\int_0^\ell dx a (1-\ps),
\label{cb3}
\end{equation}
where we have used a requirement that $Th/\gamma -\vs b$ is
a periodic function of $x$. We can derive
the expression of $\cb$ by substituting $a(x;\fm)$ 
into (\ref{cb3}). However, this expression is very complicated. 
We now simplify this. 

Eliminating $a'$ in (\ref{cb3}) by use of (\ref{abar}) and (\ref{ca4}),
we obtain
\begin{equation}
\cb=\frac{\vs}{\ell}\int_0^\ell dx (x-H)
+\frac{\vs}{\ell}\int_0^\ell dx \ps a-\vs K_1.
\end{equation}
We can simplify this equation  as 
\begin{equation}
\cb=\frac{\vs}{\ell}\int_0^\ell dx \ps \bar a-\vs K_1,
\label{33}
\end{equation}
where we have used the equality
\begin{equation}
\int_0^\ell dx (x-H)(1-\ps)=0.
\end{equation}
Substituting  (\ref{abar2}) into (\ref{33}), 
we obtain 
\begin{eqnarray}
\cb&=& \frac{\vs}{\ell (1-\e^{-\beta \fm \ell}) }\int_0^\ell dx 
\e^{-\beta (U(x)-\fm x)} \nonumber \\
&\phantom{=}&
\left[\Phi(\ell;\fm)-(1-\e^{-\beta \fm \ell})\Phi(x;\fm) \right].
\label{cbr}
\end{eqnarray}
Here we have an identity  for an arbitrary periodic 
function $\phi(x)$  with the period $\ell$:
\begin{eqnarray}
& \phantom{=}&
\int_0^\ell dy \phi(y) \e^{-\beta \fm y}-
(1-\e^{-\beta \fm  \ell})\int_0^x dy \phi(y) \e^{-\beta \fm y} \nonumber \\
&=& \e^{-\beta \fm x}\int_0^\ell dy \phi(y+x) \e^{-\beta \fm y}.
\label{ident}
\end{eqnarray}
Putting $\phi=\ps^2 \e^{\beta U}$ in (\ref{ident}), we simplify
(\ref{cbr}) as
\begin{eqnarray}
\cb&=&
\frac{\vs}{ (1-\e^{-\beta \fm \ell}) \ell}\int_0^\ell dx \e^{-\beta U(x)}
\nonumber \\
& \times &
\int_0^\ell dy
(\ps(y+x;\fm))^2e^{\beta U(y+x)-\beta \fm y}.
\label{cbr2}
\end{eqnarray}
Using (\ref{ps}) and (\ref{cur}), we can  derive
\begin{equation}
\cb=D(\fm)
\label{cbr4}
\end{equation}
with (\ref{dif}).

Finally, defining 
\begin{equation}
g(x;\fm)=c'(x;\fm)\ps(x;\fm),
\end{equation}
we rewrite (\ref{cc}) as
\begin{equation}
\frac{T}{\gamma}g^{\prime}-\vs c^\prime 
-\der{\vs}{\fm}+\frac{T}{\gamma}\pder{\ps'}{\fm}
-\ps \cc =0.
\label{cc2}
\end{equation}
Integrating this equation over the region $[0, \ell]$, we obtain
\begin{equation}
\cc =  -\der{\vs}{\fm},
\label{cc2c}
\end{equation}
where we have used a requirement that $Tg/\gamma -\vs c$ is
a periodic function of $x$.

\section{Discussion}

In conclusion, we have proposed the method of measuring 
the effective temperature of the Langevin system  (\ref{lan}) by using
a slowly varying potential and have found that this effective 
temperature is equal to $\Theta$ defined by (\ref{fdt2}).
The independence of measurements of the quantities $D$, $\mud$ and 
$\Theta$ makes  us interpret $D=\mud\Theta$ as the extended Einstein 
relation of the Langevin equation. This significant result was obtained 
by analyzing the Fokker-Planck equation with a slowly varying potential.

At the end of this paper, we shall present remarks on (\ref{FP2})
which is derived by the  analysis of the system with the moving 
potential.  (\ref{FP2}) provides  two important insights in addition 
to the direct measurement method of $\Theta$.  

The first insight obtained from (\ref{FP2}) is related 
to the interpretation  of the extended Einstein relation 
$D=\mud\Theta$ among independent 
measurable quantities.  Because (\ref{FP2}) is 
identical in form to equations that describe equilibrium systems, 
$D=\Theta/\Gamma$  (the Einstein relation in the linear response theory)  
should hold. From (\ref{gdef}),
this Einstein relation yields  $D=\mud\Theta$.  
As is well known, the Einstein relation is closely connected to the 
existence of detailed  balance for fluctuations. However, 
note that fluctuations described by (\ref{lan}) with $f \not = 0$ 
do not satisfy the detailed balance condition, as can be easily 
checked by using the steady state density (\ref{ps}). 
Therefore, we  find that the detailed balance condition is recovered through 
the coarse-graining procedure yielding (\ref{FP2}) 
and that (\ref{fdt2}) can be understood as the result of 
the recovery of  detailed balance with respect to 
the canonical distribution for the temperature $\Theta$.

The second  insight obtained from (\ref{FP2}) is related to the extension
of thermodynamics to NESS.  
When we assume that $\tilde V(Y)$ takes a tanh-like form with 
amplitude $\Delta$ (that is,  $\tilde V(\infty)-\tilde V(-\infty)=\Delta$),
we  have
\begin{equation}
\frac{Q_+-Q_-}{Q_-}=-\frac{\Delta}{\Theta}
+O \left( \left( \frac{\Delta}{\Theta} \right)^2 \right)
\label{res}
\end{equation}
with  $Q_{\pm} \equiv \lim_{Y\rightarrow\pm\infty}Q(Y)$. Then, 
the chemical potential extended to NESS can be defined 
using (\ref{res}) in a similar way as the case of
driven lattice gas \cite{HSI}. Therefore, it may be possible to
incorporate the idea of the effective temperature into 
a theoretical framework of thermodynamic functions extended to NESS.
A  study with this aim treating a wide class of nonequilibrium systems,
including many body-systems, is now in progress.

\section*{Acknowledgment}

The authors acknowledge T. Harada for stimulating discussions on 
NESS of a small bead and for a detailed explanation of experiments on
such systems. 
This work was supported by a grant from the Ministry of Education, 
Science, Sports and Culture of Japan (No. 14654064).


\begin{thebibliography}{10}

\bibitem[*]{email}
Electronic address:
hayashi@jiro.c.u-tokyo.ac.jp,\\
sasa@jiro.c.u-tokyo.ac.jp


\bibitem{FT} D. J. Evans, E. G. D. Cohen, and G. P. Morris, 
Phys. Rev. Lett. {\bf 71}, 2401 (1993).

\bibitem{jar} C. Jarzynski, Phys. Rev. Lett. {\bf 78}, 2690, (1997).  

\bibitem{ftex} G. M. Wang et al, Phys. Rev. Lett. {\bf 89}, 050601, (2002).  

\bibitem{jarex} F. Ritort, C. Bustamante and I. Tinoca, Jr.,  
 Proc. Natl. Acad. Sci. {\bf 99}, 13544, (2002).

\bibitem{hara} T. Harada and K. Yoshikawa, preprint.


\bibitem{tpp} See, e.g., introduction of Ref. \cite{RH}.

\bibitem{est} This value represent the ratio of the mass 
$\sim 10^{-15}$ g to the resistance constant $\sim 10^{-6}$ g/sec.


\bibitem{Risken}  See e.g. H. Risken, \begin{em} The Fokker-Planck Equation 
\end{em} (Springer, Berlin, 1984).  

\bibitem{RH} P. Reimann et al, Phys. Rev. Lett. {\bf 87}, 010602, (2001).     


\bibitem{CM} G. Costantini and F. Marchesoni, Europhys. Lett {\bf 48}, 491,
(1999).

\bibitem{Gall} G. Gallavotti, Phys. Rev. Lett., {\bf 77}, 4334 (1996).


\bibitem{CKP} L. F. Cugliandolo, J. Kurchan and L. Peliti, Phys. Rev. E 
{\bf 55}, 3898, (1997). 

\bibitem{BB} L. Berthier and J-L. Barrat, Phys. Rev. Lett., {\bf 89}, 
095702, (2002). 

\bibitem{jamm} I. K.  Ono et al, Phys. Rev. Lett., {\bf 89}, 095703, (2002). 

\bibitem{DVL} A. B. Kolton  et al,  Phys. Rev. Lett., {\bf 89}, 
227001, (2002). 

\bibitem{HSI} K. Hayashi and S. Sasa, Phys. Rev. E {\bf 68}, 035104(R), 
(2003). 

\end{thebibliography}
\end{document}